\documentclass{article}
\usepackage[utf8]{inputenc}
\usepackage{authblk}

\title{den}

\usepackage{graphicx}
\usepackage{hyperref}

\begin{document}

\title{A robust and non-parametric model for prediction of dengue incidence}
\author[1]{Atlanta Chakraborty}
\author[2]{Vijay Chandru}

\affil[1]{Department of Mathematics, Indian Institute of Science,   \href{mailto:atlantac96@gmail.com}{atlantac96@gmail.com}}
\affil[2]{  Center For BioSystems Science And Engineering,
 Indian Institute of Science, \href{mailto:chandru@strandls.com}{chandru@strandls.com}
 }
\date{}

\maketitle

\begin{abstract}
Disease surveillance is essential not only for the prior detection of outbreaks but also for monitoring trends of the disease in the long run. In this paper, we aim to build a tactical model for the surveillance of dengue, in particular. Most existing models for dengue prediction exploit its known relationships between climate and sociodemographic factors with the incidence counts, however they are not flexible enough to capture the steep and sudden rise and fall of the incidence counts. This has been the motivation for the methodology used in our paper. We build a non-parametric, flexible, Gaussian Process (GP) regression model that relies on past dengue incidence counts and climate covariates, and show that the GP model performs accurately, in comparison with the other existing methodologies, thus proving to be a good tactical and robust model for health authorities to plan their course of action.

\end{abstract}

\section{Introduction}
\label{intro}
Dengue is a fast emerging pandemic-prone viral disease transmitted by \textit{Aedes aegypti} and \textit{Aedes albopictus} mosquitos. According to the World Health Organisation (WHO), each year, an estimated 390 million dengue infections occur all around the world. Cases across the Americas, South-East Asia and Western Pacific exceeded 1.2 million in 2008 and over 3.2 million in 2015 \cite{url1}. Several precautionary measures include vector control tools, like controlling mosquito populations, however, implementation is a major challenge and effective dengue prevention is rarely achieved, specially in developing countries. Often, it is the emergency vector control operations that is usually applied when an outbreak occurs, like insecticide fogging. 

Accurate forecasts of incidence cases, or infected individuals are key to planning and resource allocation of dengue vaccines, medical centres, etc. Previous attempts to model dengue have made use of relatively simple models, such as generalised linear model and ARIMA, exploiting the relationship with other environmental variables (\cite{climate1}, \cite{climate2}, \cite{timeclimate}). However, most of the times, disease dynamics are not well understood and such models may fail to capture that \cite{failcapture}. Dengue is closely related to the seasonal changes, rainfall and humidity. Our model is trained on historical incidence data, mean surface temperature, humidity and rainfall, and makes use of a Bayesian non-parametric modelling framework, Gaussian processes (GP), that allows for flexibility in the model, thus being able to forecast the sudden peak increase of the incidence counts. 

\medskip

\section{Related Work}

A study and systematic review of existing dengue modelling methods, conducted by Louis et al, \cite{failcapture}, has been instrumental in providing us with an overview of current modelling efforts and their limitations. The study enlists a wide range of predictors that were used to create dengue risk maps, like socioeconomic and demographic data, climatic and environmental data, remote sensing and entomological data. 

Several efforts (\cite{climate2}, \cite{climate1}, \cite{timeclimate}, \cite{dengueg}) have made use of parametric models like logistic regression models, multinomial models and generalised linear models with climatic covariates as inputs. Climatic data have been found to be particularly useful for the generation of risk maps \cite{gaussdengue}. Additionally, other factors, such as human mobility or housing conditions, are also likely to be linked to the occurrence of dengue cases \cite{11}. In contrast to the fields of malaria and other vector borne diseases, the study shows that dengue is particularly challenging due to the high number of non-detectable breeding sites (\cite{13}, \cite{14}, \cite{15}, \cite{16}). There have been models making use of entomological data (\cite{3}, \cite{5}, \cite{10}, \cite{15}, \cite{18}, \cite{26}, \cite{25}) studying the link between several vector aspects (like larvae abundance, ovi-trapping) and dengue cases, however the exact nature of the relationship remains unknown. Such surveys are not only labor-intensive and costly, but they also yield spurious results that are not useful for prediction (\cite{5}). 

The weakness of current dengue prediction efforts originates from the fact that dengue is highly dynamic and multifactorial. Factors like host immunity, genetic diversity of circulating viruses, etc also play an important role, but they are difficult and expensive to track. They continue to pose challenges and limit the ability to produce accurate and effective risk maps and models, thus failing to support the development of an early warning system.  Many epidemiological models have developed and have gained importance in the last decade, however, they cannot be used in the public health context due to their complexity and the extensive need for input data. Additionally, most models produce an average forecast of the numbers in the long run, instead of being able to predict the immediate rise and fall of the incidence counts. On account of this, we aim to build a robust but easily implementable model that would depend only on climatic variables and past historical data. The GP model has an added advantage of generalising the model to include several other factors that affects dengue, like human mobility, vector data, etc.

\medskip

\section{The data}

We have chosen Singapore as our case study due to the free availability of weekly dengue incidence counts. Data for the years 2005-2017 was downloaded from \cite{url2}. Rainfall, relative humidity and surface air temperature data was also downloaded from NEA website. 2005-2016 have been used as a training set, whereas the year 2017 was used as a test set. The incidence data are final counts, i.e. the total number of cases in each week. In order to avoid overfitting or under-fitting, we have implemented the process of k-fold cross validation (k=10) to help us understand the performance of the model, and select an appropriate one.

\begin{figure*}
 \includegraphics[width=180mm]{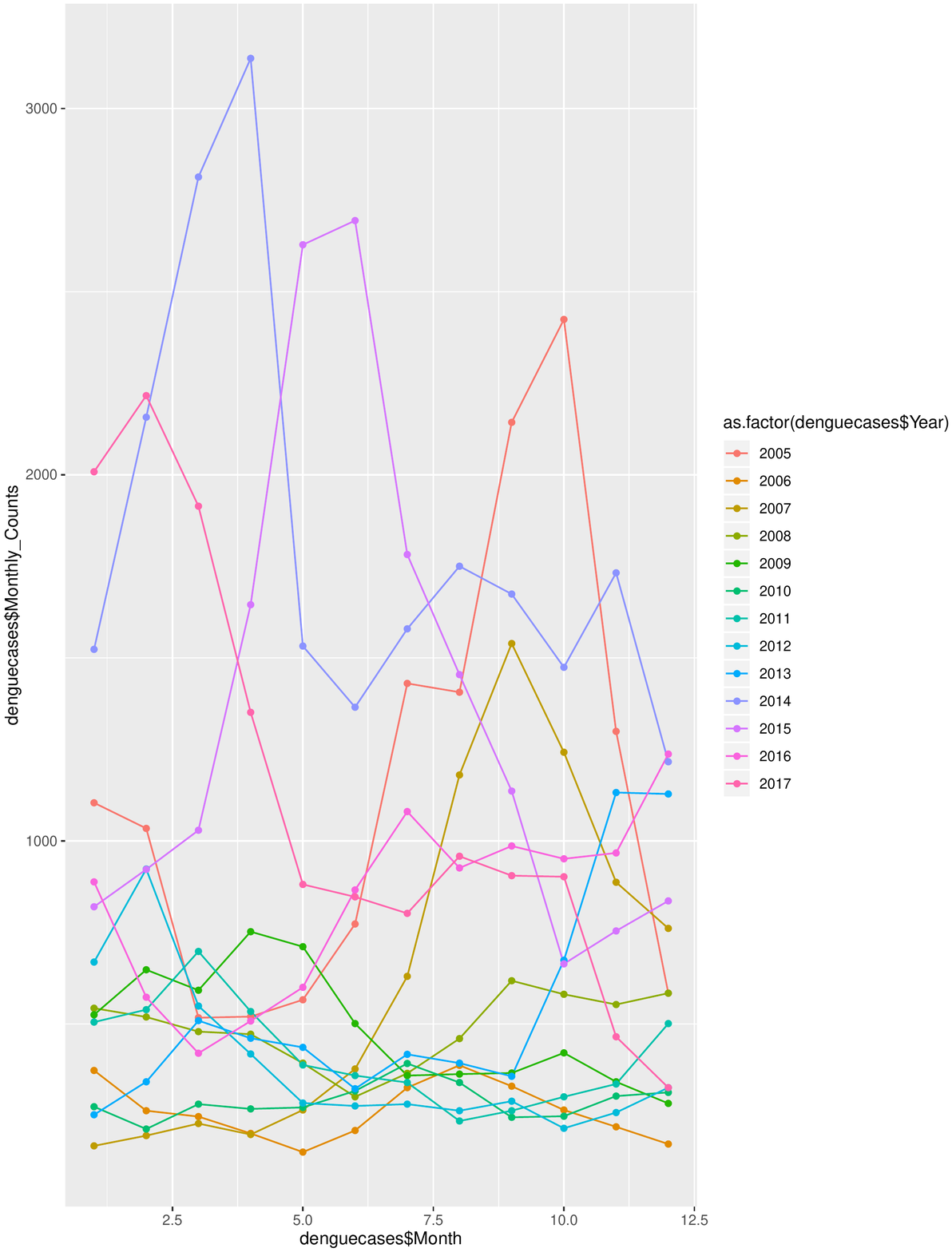}
\caption{Monthly incidence of dengue vs month across years 2005-2017}
\label{fig: timeseries}
\end{figure*}

Figure \ref{fig: timeseries} shows Singapore incidence counts across 2005-2017. It clearly depicts a yearly circle. Not only does the data have a mean response which varies with time, but also the variability of the incidence counts is unequal across the months. The dengue count is a heteroskedastic variable when predicted by the month number. 

\medskip

\section{Methodology}

\subsection{Gaussian process modelling}

This paper proposes to model dengue incidence with Gaussian processes (GP), a non-parametric modelling framework \cite{gausbook}, for the purpose of getting an added flexibility and making accurate predictions of the peak season, as it falls and rises. One can think of GP as defining a distribution over functions, and inference takes place directly in the space of functions. 

A GP is completely specified by its mean function $m(\mathbf{x})$ and the covariance function $k(\mathbf{x},\mathbf{x'})$ of a process $f(\mathbf{x}),$ and we write $f(\mathbf{x}) \sim \mathcal{GP} (m(\mathbf{x}), k(\mathbf{x},\mathbf{x'})).$ We first assume our model to be of the form $y=f(\mathbf{x})+\epsilon,$ where $\epsilon$ is additive and independent identically distributed Gaussian noise with variance $\sigma_n^2.$

From our training data, we know $\{(\mathbf{x_i}, y_i)|i=1,...,n \}$, where $n$ is the total number of observations. The joint distribution of the training outputs, $y$, and the test outputs $f_*$ according to the prior is 
$$\left[{\begin{array}{c} y \\
    f_* \end{array}}\right]  \sim \mathcal{N} \left(0, \left[{\begin{array}{cc} K(\mathbf{X},\mathbf{X})+\sigma_n^* \mathbf{I} & K( \mathbf{X},\mathbf{X_*} ) \\
 
    K(\mathbf{X_*}, \mathbf{X}) & K(\mathbf{X_*}, \mathbf{X_*}) \end{array}}\right] \right).$$
If there are $n_*$ test points then $K(\mathbf{X}, \mathbf{X_*})$ denotes the $n \times n_* $ matrix of the covariances evaluated at all pairs of training ($X$) and test points ($X^*$), and similarly for the other entries. To get the posterior distribution over functions we need to restrict this joint prior distribution to contain only those functions which agree with the observed data points. The key predictive equations for Gaussian process regression are
\begin{equation}
f_*|X, y, X_* \sim \mathcal{N} (\bar{f_*}, cov(f_*))  
\end{equation}
\begin{equation}
\bar{f_*}= K(X_*, X){[K(X,X)+\sigma_n^2 \mathbf{I}]}^{-1}y 
\end{equation}
\begin{equation}
cov(f_*)= K(X_*, X_*)- K(X_*, X){[K(X,X)+\sigma_n^2 \mathbf{I}]}^{-1} K(X, X_*).
\end{equation}

By this definition, GPs allow us to obtain the exact predictive distribution through a closed-form expression. They are also flexible, since one can use any positive semi-definite kernel as the covariance function $K$ as a measure of similarity between points, providing rich insights about the dependencies between them. 

Under the Gaussian process model, the prior is Gaussian, $f|X \sim  \mathcal{N} (0,K),$ or 
\begin{equation}
\log p(f|X)= \frac{-1}{2} f^T K^{-1}f - \frac{1}{2} \log |K| -\frac{n}{2} \log 2 \pi 
\end{equation}
and the likelihood is a factorised Gaussian $y|f \sim \mathcal{N}(f, \sigma_n^2 \mathbf{I}).$ We thus arrive at the log marginal likelihood as 
\begin{equation}
 \log p(y|X)= \frac{-1}{2} y^T (K+\sigma_n^2 \mathbf{I})^{-1}y - \frac{1}{2} \log |K+\sigma_n^2 \mathbf{I}| -\frac{n}{2} \log 2 \pi .
\end{equation}

We estimate the hyper-parameters of $K$ by maximising the marginal likelihood (or minimising the negative log likelihood). We can use several gradient-based optimisers, since it is necessary to compute the partial derivatives of the marginal likelihood w.r.t. the hyper-parameters. For our purpose, we use the ``BFGS" method. 

We apply a logarithmic (one plus) transformation on the response variable and model this transformation as a GP. This is done to ensure that the largest variances are stabilised. The main task in modelling via GPs is to define an appropriate covariance structure. We assume a zero-mean GP by centering the response variable about its mean. 
	
\medskip

\subsection{Defining the Covariance function}

Covariance functions encode our assumptions about the function which we wish to learn. It is a basic assumption that input points which are ``close" to each other are likely to have similar target values $y$. Based on this, training points that are close to a test point should provide information about the prediction at that point. It is the covariance function that defines this nearness or similarity. 

A complex covariance function is derived by combining several different kinds of simple covariance functions. The covariance structure imposed by the GP prior should reflect what we expect from the data. We make use of standard kernels defined in the GP literature \cite{gausbook}. Our goal is to model the transformed incidence counts as a function of $x_i= (x_1, x_2, x_3, x_4)_i,$ i.e. the $i$th observation and its corresponding month number, total monthly rainfall, mean relative humidity and mean surface air temperature respectively.

To enforce the assumption that the test input is highly correlated with its pre-ceding inputs, we use a 5/2 Matern Kernel which is defined as 
\begin{equation}
k_1(x_i, x_j)= \sigma_{1}^2 \left( 1+ \frac{\sqrt{5} \Delta x}{l_{1}} +\frac{5\Delta x^2}{3l_{1}^2} \right) exp \left( \frac{-\sqrt{5} \Delta x}{l_{1}}\right) 
\end{equation}
where $\Delta x=|x_i-x_j|$ is the absolute distance between the inputs. Its hyper-parameters, $\sigma_{1}$ and $l_{1}$ are used to control the strength of correlation signal and the span of time that should correlate, respectively. 

We use a second component to exploit the periodicity observed in dengue incidence, while still giving more importance to closer periods of time. 
\begin{equation}
k_2(x_i, x_j)= k_{21} (x_i, x_j)\times k_{22} (x_i, x_j) 
\end{equation}
\begin{equation}
k_{21}(x_i, x_j)= \sigma_2^2 exp \left( \frac{- \Delta x^2}{2 l_2^2} 
\right) 
\end{equation}
\begin{equation}
k_{22} (x_i, x_j)= exp \left( -2 sin^2 \left( \frac{\pi \Delta x}{p} \right) / l_{per}^2  \right)
\end{equation}
$k_{21}$ is a squared-exponential kernel (also called radial basis function kernel) and $k_{22}$ is a periodic kernel. The hyper-parameters of $k_{21}$- $l_{2}$ and $\sigma_{2}$ are used to control the number of months that should impact the incidence and strength of the correlation signal respectively. $p$ and $l_{per}$, the hyper-parameters of $k_{22}$ are used to control the periodicity and length-scale of the signal respectively.
 
Next, we model the small irregularities with a rational quadratic term. The rational quadratic kernel allows us to model the data varying at multiple scales. 
\begin{equation} 
k_3 (x_i, x_j)= \sigma_3\left( 1+\frac{\Delta x^2}{2 \alpha l_3^2} \right)^{- \alpha}
\end{equation}
$\sigma_3$ is the magnitude, $\alpha>0$ is the scale-parameter and $l_3$ is the characteristic length-scale.

Finally, we specify the noise model as the sum of a squared exponential contribution and an independent component. Noise in the series could be due to measurement inaccuracies. It could also be due to the changes in weather phenomena every year, hence we assume that there is a little amount of correlation in time.   
\begin{equation}
k_4(x_i, x_j)=   \sigma_f^2 exp \left( \frac{- \Delta x^2}{2 l_4^2} \right)+ \sigma_n ^2 \delta_{x_i x_j}
\end{equation}
where $\sigma_f$ is the signal variance, $l_4$ is its length scale and $\sigma_n$ is the magnitude of the independent noise component. 

The final covariance function is
\begin{equation}
k(x_i, x_j)= k_1(x_i, x_j)+ k_2(x_i, x_j)+ k_3(x_i, x_j)+k_4(x_i, x_j)
\end{equation}
with 12 hyper-parameters. 

Note that most of the above defined covariance functions are stationary, i.e. invariant to translations in the input space. Sampson and Guttorp, \cite{warping}, introduced the method of warping in 1992, which allows us to introduce an arbitrary non linear map $u(x)$ of the input space $x$, and then use stationary covariance functions in the $u(x)$ space. This is yet another reason for the logarithmic transformation of the incidence counts.  

The code has been run on R, mainly using the \texttt{GauPro} package. All the above mentioned kernels are imported from the \texttt{kergp} package. The training data is then fit and the marginal likelihood is optimised using the ``BFGS" algorithm. 

\medskip

\section{Results and Discussion}

\begin{figure*}
\includegraphics[width=122mm]{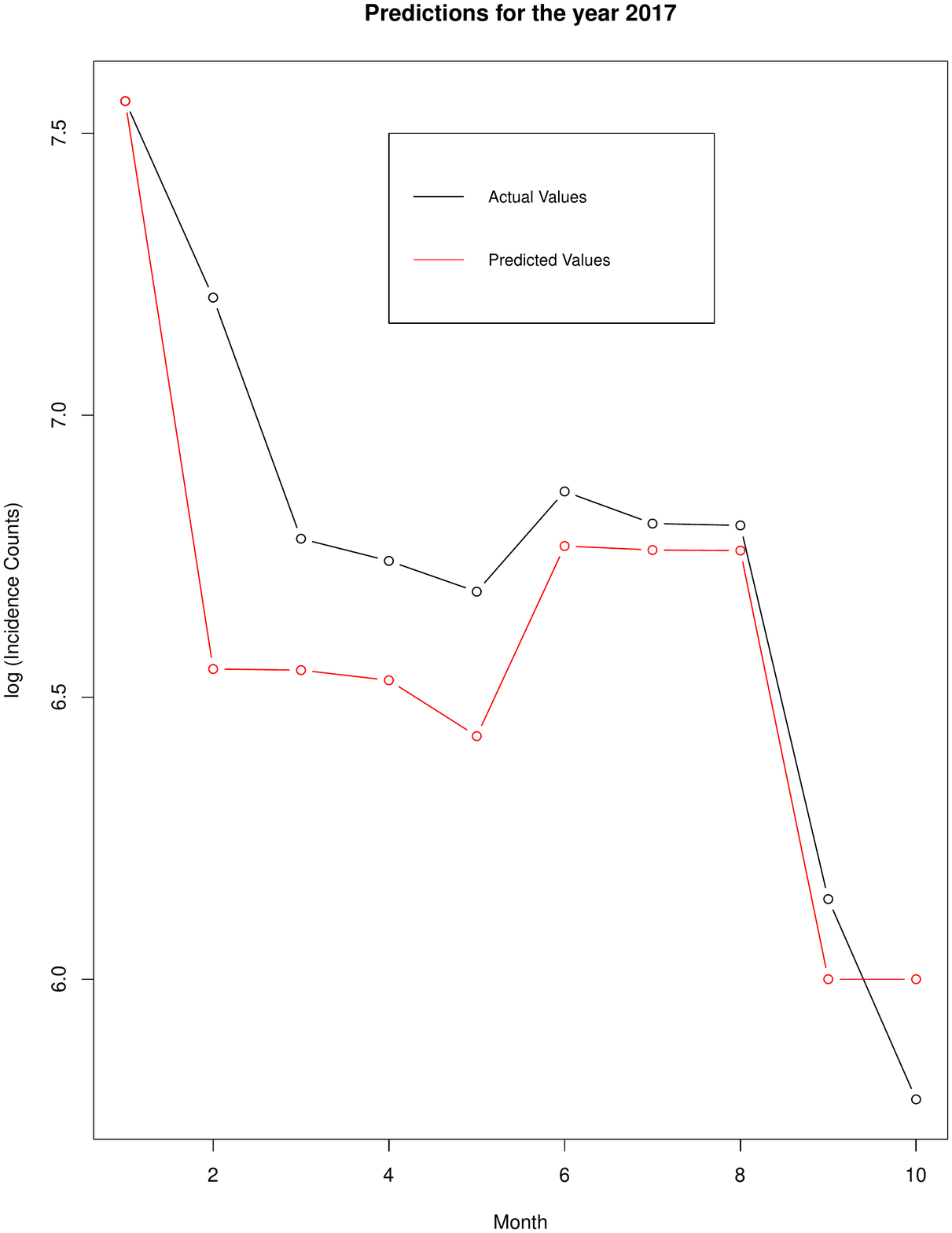}
\caption{Gaussian model on the test set}
\label{fig: test}
\end{figure*}

We compare our GP model with 3 different existing methodologies- time series forecasting (Arima), generalised additive models (GAM) and predictions from random forests (RF), on the basis of two different metrics- root mean squared error (RMSE) and mean absolute deviation (MAD). The performance across various methods is reported as follows, for both in sample and out sample forecasts.
\begin{center}
 \begin{tabular}{||c || c | c || c | c ||} 
 \hline
Model & Training RSME & Test RMSE & Training MAD & Test MAD \\ [0.5ex] 
 \hline\hline
GAM & 0.608 & 0.682 & 0.623 & 0.883\\
 \hline
 Time series & 0.521 & 0.562 & 0.493 & 0.512\\
 \hline
 Random Forest & 0.676 & 0.719 & 0.713 & 1.101\\
 \hline
 GP & 8.3e-07 & 0.260 & 9.63e-07 & 0.262\\
 \hline
\end{tabular}
\end{center}
Overfitting is taken care of by cross validation. As can be seen from the above performance metrics, GP is very accurate for forecasting and easily implementable. It also has room for adding more covariates to the model. For the out-of-sample forecasts, Figure \ref{fig: test} shows the predictions for the year 2017. 
\medskip

\section{Conclusion and Future Work}
As we have seen in the last section, the model fit by Gaussian process serves as a good tactical model. This is due to its non-parametric nature and its flexibility, thus being able to automatically adapt to different scenarios. The other advantage of this model is the nature of the input, incidence numbers correlated with climate variables. In the future, we would like to investigate the role of human and vector factors in helping us forecast dengue incidence in a public health context. The GP model can accommodate such factors by introducing kernel functions based on human and vector interactions and add it to the already defined kernel function in this paper.

The GP model provides a sufficient window for health authorities to be aware of the incoming dengue counts and hence carefully plan and take necessary actions.  Its easy implementation can act as a very accurate early warning system when implemented on a weekly basis.To make the model functional on a weekly basis, one may consider data on a weekly scale spatially and consider resource allocation and facility problems to effectively implement an operational model.

\medskip

\section{DECLARATIONS}
\begin{enumerate}
    \item Ethics approval and consent to participate- not applicable
    \item Consent for publication- not applicable
    \item Availability of data and material- Data publicly available on Ministry    of    Health    Singapore:https://www.moh.gov.sg/resources-statistics/infectious-disease-statistics/2018/weekly-infectious-diseases-bulletin, (2018)
    \item Competing interests- The authors declare that they have no competing interests.
    \item Funding- No funding received
    \item Authors' contributions- All authors contributed equally, read and approved the final manuscript.
    \item Acknowledgements- Not applicable
    \item Authors' information (optional)
\end{enumerate}

\end{document}